\documentclass[twocolumn,aps,prl,nofootinbib,superscriptaddress,showkeys]{revtex4-1}

\usepackage[colorlinks=false,linkcolor=black,citecolor=black]{hyperref}
\usepackage{graphicx}
\usepackage{subfigure}
\usepackage{amstext,amsmath,amsfonts,amssymb,amsthm}
\usepackage[format=plain,justification=centerlast]{caption}
\usepackage{hyperref} 
\usepackage{color}

\usepackage{verbatim}

\def\eq#1{{(\ref{#1})}}

\def\be{\begin{equation}}
\def\ee{\end{equation}}
\def\bes{\begin{eqnarray}}
\def\ees{\end{eqnarray}}
\def\ba{\begin{align}}
\def\ea{\end{align}}
\def\bwt{\begin{widetext}}
\def\ewt{\end{widetext}}

\def\br{\bar{r}}

\def\Oo{\Omega}

\def\f{\frac}

\def\nn{\nonumber}

\begin{document}

\title{On the Penrose inequality in anti-deSitter space}
\date{\today}

\author{Viqar Husain}
\email{vhusain@unb.ca}
\affiliation{Department of Mathematics \& Statistics, University of New Brunswick, Fredericton E3B 5A3 Canada.}

\author{Suprit Singh}
\email{suprit.singh@unb.ca; supritsingh@gmail.com}
\affiliation{Department of Mathematics \& Statistics, University of New Brunswick, Fredericton E3B 5A3 Canada.}

\begin{abstract}

For asymptotically flat spacetimes the Penrose inequality gives an initial data test for the weak cosmic censorship hypothesis. We give a formulation of this inequality for asymptotically anti-deSitter (AAdS) spacetimes, and show that the inequality holds for time asymmetric data in spherical symmetry. Our analysis is motivated by the constant-negative-spatial-curvature form of the AdS black hole metric.

\end{abstract}


\maketitle

 One of the most important unresolved questions in classical general relativity is Penrose's cosmic censorship hypothesis. The weak version of this hypothesis, Weak Cosmic Censorship (WCC), is the statement that a spacetime singularity arising from gravitational collapse of ``reasonable" matter must be shrouded by an event horizon \cite{Penrose:1969pc}.  

 Penrose also proposed an ``initial data test"   suggested by the WCC hypothesis. Consider an asymptotically flat solution of Einstein equations  with matter satisfying the dominant energy  condition. Then if a Cauchy slice  of this solution contains an outer-trapped  2-surface $S$ of area $A(S)$, and if $M$ is the Arnowitt-Deser-Misner (ADM) mass of the data on the slice, the conjectured inequality is 
\be
A(S) \leq  16\pi M^2.
\ee
  The intuition behind this inequality is that the mass $m_{ah}$ contained within the outermost  apparent horizon  cannot exceed the total mass $M$ of the data. This is expected to be the case for an ongoing gravitational collapse since, in general,  the region between this  horizon and spatial infinity will contain uncollapsed (positive energy) matter.  Thus the inequality may equivalently be written as
\be
m_{ah} \le M, \label{ineq-flat}
\ee 
with equality in the quiescent black hole state, when all matter has fallen into the horizon. A review of the status of WCC and the Penrose inequality appears in Refs. \cite{Wald:1997wa,Mars:2009cj} 

In this paper we  study a generalization of this inequality for asymptotically anti-deSitter (AAdS) spacetimes in 4-spacetime dimensions. This requires definitions of AAdS data and quasi-local mass. The former is well known (see eg. \cite{Henneaux:1985tv}),  and the latter requires a suitable generalization of one several mass definitions \cite{Szabados:2009eka} to AAdS. The basic requirement is that any quasi-local mass should give the total mass of data in the asymptotic limit. This is guided by, and tested for, using the Schwarzschild-AdS spacetime. 

We will use the Hawking mass \cite{Hawking:1968qt}, which has been used to study the Penrose inequality in the asymptotically flat case 
\cite{Malec:2002ki}. This quantity is a measure of the mass contained within a closed 2-surface $S$ in a spatial slice of spacetime. The definition is   
\be
m(S) \equiv  \sqrt{\frac{A_S}{16\pi}} \left[ 1+ \frac{1}{16\pi} \int_S \theta_+\theta_- \right], \label{h-mass}
\ee
where   $A_S$ is the area of $S$ and  $\theta_\pm$ are the future directed ingoing ($-$) and outgoing ($+)$ null expansions of $S$. In terms of the ADM data $(q_{ab},K_{ab})$ these are
\be
\theta_\pm = (q_{ab} - s_as_b)(K_{ab}\pm D_as_b ) 
\ee 
where $s^a$ is the unit normal to $S$.
 
As a prelude to introducing the class of data we will work with, let us note that the AdS black hole may be written in a form analogous to the Painleve-Gulstrand (PG) form, but rather than using the flat-slice PG form for AdS, it is more convenient to work with $3-$slices of constant negative curvature. With such a choice of coordinates,  the AdS black hole metric is  
\bes
ds^2 &=& -\left(1-\f{2M}{r}+\f{r^2}{l^2}\right) dt^2+2\sqrt{\f{2M}{r(1+ r^2/l^2)}} dt dr \nn\\
&& + \f{dr^2}{(1+r^2/l^2)}+r^2d\Oo^2.\label{ads-bh}
\ees
It is readily verified that this is a solution of the vacuum Einstein equations with cosmological constant $\Lambda = -3/l^2$. It is not one of the commonly used forms, (we have not seen it in the literature), but  it is a convenient choice  for our analysis; the corresponding general time dependent form (see \eq{gen-metric} below) may be useful for other applications with matter coupling.

Let us first compute the Hawking mass for the AdS black hole by choosing the $2-$surface $S$  be a spherical surface of radius $r$ in a $t=$ constant slice  of this metric. A straightforward calculation  gives 
\be
\theta_+\theta_- = \frac{8M}{r^3} - \frac{4}{r^2} - \frac{4}{l^2}, 
\ee 
and 
\be
m(r) = M- \frac{r}{2} \left( \frac{r}{l} \right)^2.
\ee
In the $r\rightarrow \infty$ limit this gives a divergent result, where we should get $M$. This problem is rectified by the ``background subtraction" given by  
\be
\tilde{m}(S) \equiv  \sqrt{\frac{A}{16\pi}} \left[ 1+ \frac{1}{16\pi} \int_S \left( \theta_+\theta_- + \frac{4}{l^2} \right)  \right]. 
\label{mH}
\ee   
This modified Hawking  mass gives the expected result  for the above metric, namely
$\displaystyle \lim_{r\to \infty} \tilde{m}(r) = M$. 

From the definition \eq{mH}, we note also that the Hawking mass contained within a marginally outer-trapped surface $S_{ah}$ (i.e. a surface on which $\theta_+=0$) is 
\be
m_{ah}\equiv\tilde{m}(S_{ah})= \sqrt{\frac{A_{ah}}{16\pi}} \left[ 1+ \frac{A_{ah}}{16\pi}  \left( \frac{4}{l^2} \right)  \right], \label{mah}
\ee 
where $A_{ah}$ is the area of $S_{ah}$.
Therefore a natural proposal  for the Penrose inequality for AAdS data, motivated by  (\ref{ineq-flat}),   is 
\be
\tilde{m}(S_{ah})  \le M,\label{ineq-ads}
\ee 
where again $M$ is the total mass of the data; $M$ may be computed using any asymptotic mass formula (see eg. \cite{Balasubramanian:1999re,Ashtekar:1999jx}),  and must agree with that obtained from the asymptotic limit of \eq{mH}.

There are proposals for the Penrose inequality similar to \eq{ineq-ads} for AdS spacetimes which use a background subtraction to define a regulated mass. One of these is an extension to include charge  \cite{Itkin:2011ph}, but contains no further analysis. Another work 
\cite{Cha:2017gej} presents a proof of a simpler inequality, $M\ge \sqrt{A_{ah}/16\pi}$, which excludes the second term on the r.h.s. in  \eq{mah}. Similar inequalities in AdS space have studied in related contexts. Ref. \cite{Bengtsson:2016dac} contains a formulation and proof in 2+1 AdS spacetime; Ref. \cite{Fischetti:2016vfq} studies the generalized Hawking mass in a holographic context using the technique of inverse mean curvature (IMC) flow (see also \cite{Gibbons:1998zr} where IMC flow with Hawking mass is suggested as a means for proving Penrose inequalities). 

Our proof of \eq{ineq-ads} does not use IMC flow, instead it uses instead the flow generated directly by the constraint equations with matter sources. This gives a radial equation  for the modified Hawking mass. 

The first step is a construction of  AAdS data $(\pi^{ab}, q_{ab})$ that is a solution of the Hamiltonian and diffeomorphism constraints 
\bes
&&\left(\pi^{ab}\pi_{ab} - \frac{1}{2} \pi^2  \right) - R(q)  - \frac{6}{l^2}  +\rho_M = 0,\\
&& -2D_b\pi^{ab}= j_M^a,
\ees
where $(\rho_M,j_M)$ is matter density and current, and $\pi=\pi^{ab}q_{ab}$.  Specifically let us take the spherically symmetric, and  time asymmetric data  
\bes
\pi^{ab} &=& f(r)s^as^b + g(r)q^{ab}, \nn\\
q_{ab}dx^adx^b &=&   \f{dr^2}{(1+r^2/l^2)} + r^2 d\Oo^2, \label{data}
\ees  
where $s^a$ is  the unit radial vector. 

This form is guided by the metric (\ref{ads-bh}), and  represents  general spherically symmetric data (but excludes spatial metric wormholes akin to Misner data \cite{Misner:1960zz}).  To see this let us recall that the corresponding phase space, before gauge fixing, is specified by two spatial metric functions,   $ds^2_q=\lambda^2(r) dr^2 + \sigma(r)^2d\Omega^2$, and their conjugate momenta which appear in the ADM momentum $\pi^{ab}$.  The gauge choice $\sigma(r)=r$, and $\lambda(r)$ as specified in \eq{data}, leave the two functions $f$ and $g$ in our form for $\pi^{ab}$.  The choice of $\lambda(r)$ gives a 3-metric of constant negative curvature so that dynamical information is encoded entirely in $\pi^{ab}$; the corresponding spacetime metric is of the form 
\bes
ds^2 &=& -a(r,t) dt^2 + 2b(r,t) drdt  \nn\\
 && + \f{dr^2}{(1+r^2/l^2)}+r^2d\Oo^2, \label{gen-metric}
\ees
and the static vacuum solution is \eq{ads-bh}. (For a comparison with  flat slice data with zero cosmological constant see  \cite{Guven:1999hc}; the paper also comments on constant negative curvature foliations.) 
    
A useful feature of the form \eq{data} is that the Ricci scalar term exactly cancels the cosmological constant in the Hamiltonian constraint, so  the constraint equations reduce to 
\bes
&& (f-3g)(f+g) + 2\rho_M(r) =0, \label{h}\\
&& \frac{d}{dr}(f+g) + \frac{2f}{r}= -\ \frac{j_M(r)}{2(1+r^2/l^2)}.\label{diff}
\ees
The null expansions are 
\be
\theta_\pm = -(f+g) \pm \frac{2}{r} \sqrt{1+ \left(\frac{r}{l} \right)^2}, 
\ee
which, for spherical surfaces, gives the Hawking mass 
\be
\tilde{m}(r) = \frac{r^3}{8} (f+g)^2. \label{sph-mass}
\ee
The mass contained within an apparent horizon of radius $r_{ah}$  obtained from \eq{mH} with $\theta_+=0$ is
\be
\tilde{m}(r_{ah}) =  \f{r_{ah}}{2}\left[ 1+ \left( \frac{r_{ah}}{l}\right)^2\right]. \label{hmass}
\ee

A useful formula for proving the inequality \eq{ineq-ads} is the radial mass flow equation obtained by differentiating \eq{sph-mass}, and using  \eq{h} and \eq{diff}.  This gives 
\be
\tilde{m}' = \frac{1}{4} \rho_M r^2 \left(1 \pm \frac{|j_M|}{\rho_M} \frac{1}{(1+r^2/l^2)} \sqrt{\frac{2\tilde{m}}{r} } \right), \label{jne0}
\ee
where $\pm$ on the r.h.s come from the possible signs of $j_M$.  It is convenient to reintroduce Newton's constant $G$ and work with the dimensionless variables 
\bes
\bar{m} &=& l\tilde{m}, \ \  \bar{r} = r/l, \ \  \epsilon = |j_M|/\rho_M, \nn\\
 \bar{\rho}&=&  Gl^2 \rho_M, \ \  \mu =G/l^2. \nn
\ees
Then \eq{jne0} becomes
\be
\frac{d\bar{m}}{d\bar{r}} = \frac{1}{4\mu} \bar{\rho} \bar{r}^2  \left(  1 \pm  \frac{\epsilon \sqrt{\mu}}{(1+ \bar{r}^2)} \sqrt{\frac{2\bar{m}}{\bar{r}} } \right), \label{mbar-eqn}
\ee 
and the horizon mass equation \eq{hmass} becomes  
\be
\bar{m}(\bar{r}_{ah}) = \frac{\bar{r}_{ah}}{2\mu} \left( 1+ \bar{r}_{ah}^2 \right).  \label{hmass2}
\ee 

We now proceed to prove the inequality \eq{ineq-ads} subject to the requirements that (i) any non-zero $\rho_M$ must be such that 
\be
\rho_M >0 \quad  \text{and} \quad \displaystyle \lim_{r\rightarrow\infty} \tilde{m}(r)=M, 
\ee
and (ii) the dominant energy condition (DEC), $0< \epsilon \le 1$. 

The proof for the case $j_M=0$ is immediate. Eqn. \eq{mbar-eqn} becomes 
\be
\frac{d\bar{m}}{d\bar{r}} = \frac{1}{4\mu} \bar{\rho} \bar{r}^2. 
\ee
Therefore, since $\bar{\rho}>0$,  $\bar{m}(r) $ is an increasing function on $\mathbb{R}^+$ that asymptotes to  $\bar{M}\equiv lM$ from below. Hence any horizon radius $\bar{r}_{ah}>0$ must be such that $\bar{m}(\bar{r}_{ah}) < \bar{M}$. 
 
For $j_M\ne 0$ and the positive sign in \eq{mbar-eqn}, the proof is the same:  $\bar{m}' >0$ for all $\bar{r}$, hence $\bar{m}$ has no critical points, and reaches a horizontal asymptote, corresponding to the total mass $M$, from below.  
 
For the negative sign in \eq{mbar-eqn} let us note the following two properties. 
 
 (i) $\displaystyle \frac{d\bar{m}}{d\bar{r}}|_{\bar{r}=\bar{r}_{ah}} >0$  for data such that there is an apparent horizon at some $\bar{r}=\bar{r}_{ah}$. To show this 
substitute $\bar{r} = \bar{r}_{ah}>0$ in  \eq{mbar-eqn} and use \eq{hmass2}. We then have 
\be
\bar{m}'(\bar{r}_{ah}) = \frac{1}{4\mu} \bar{\rho} \bar{r}^2_{ah} \left(  1 -  \frac{\epsilon(\bar{r}_{ah})}{\sqrt{1+ \bar{r}^2_{ah}} } \right) >0,
\ee
where the last inequality follows from the DEC,  $\epsilon(r) \le 1$.  
 
(ii) $\bar{m}(\br)$ does not have a local maximum.  To show this assume there is a local maximum at $\br_c$. Then there are close points $\br_L$ and $\br_R$, with $ \br_L<\br_c<\br_R$, such that  $\bar{m}(\br_L) = \bar{m}(\br_R) =\sigma$ and $\bar{m}'( \br_L)>0$ and $\bar{m}'( \br_R)<0$.  Then \eq{mbar-eqn} (with the negative sign we are now considering) implies 
\be
\frac{\sqrt{\mu}\ \epsilon(\br_L)  }{(1+ \br^2_L)} \sqrt{\frac{2\sigma}{\br_L} } < 1<\frac{\sqrt{\mu}\ \epsilon(\br_R) }{(1+ \br^2_R)} \sqrt{\frac{2\sigma}{\br_R} }.
\ee
This leads to a contradiction since $\br_L< \br_R$, and we can take $\epsilon(\br_L)\approx \epsilon(\br_R)$  because $\br_L$ and $\br_R$ may be taken arbitrarily close.

Properties (i) and (ii) lead to a proof  of the inequality \eq{ineq-ads}   for  our spherically symmetric AdS data. This is seen  as follows: a violation means that $\bar{m}(\br_{ah}) >\bar{M}$, where property (i) must hold.  But this means that $\bar{m}(\br_{ah})$ must go through a local maximum for some $\br > \br_{ah}$, and then ultimately cross below its horizontal asymptote at $\bar{M}$ in order to satisfy the large $r$ fall-off; (recall $m(r)$ must reach its asymptotic value from below with positive slope). But a local maximum is ruled out by property (ii).  Hence the inequality follows.


  In summary, we have given a proof of  a natural formulation  \eq{ineq-ads} of the Penrose inequality for AAdS spacetimes with   spherical symmetry.  This was done with the choice of slicing \eq{data}  in which $3-$spaces are of constant negative curvature; this corresponds to writing the general spherically symmetric metric in the form \eq{gen-metric}. Along the way, we used the constant-negative-spatial-curvature form of the AdS black hole, which together with its time dependent generalization \eq{gen-metric}, may be useful for other applications. 

Our results provide several directions for further work. Among the more important is a generalization  of our data to axial symmetry in light of recent results hinting at a violation of WCC in AAdS spacetime \cite{Crisford:2017zpi}. This work  suggests that it  may be possible to construct explicit examples that show violation of \eq{ineq-ads}. Furthermore, in light of the AdS/CFT duality conjecture, an interesting direction would be an exploration of the analog of the inequality in the CFT. This may shed light on signatures of trapped surfaces in the CFT, perhaps via properties of world-surfaces that stretch between a bulk-trapped surface and a closed $2-$ surface on the boundary spacetime. Lastly, the constant-negative-curvature spatial slice version of asymptotically AdS spacetimes suggested by \eq{gen-metric} may be useful for further numerical explorations of AdS instability \cite{Bizon:2014nhh}.
   \medskip
   
 \noindent{\it Acknowledgements} VH is supported by a Discovery Grant from the Natural Science and Engineering Research Council of Canada; SS is supported by the SERB OPDF scheme of the Government  of India.  We thank Sanjeev Seahra for comments on the manuscript. 
 
\bibliography{penroseineq}

\end{document}